\documentclass[11pt]{article}
\newcommand{\be}{\begin{equation}}
\newcommand{\ee}{\end{equation}}
\newcommand{\ba}{\begin{eqnarray}}
\newcommand{\ea}{\end{eqnarray}}
\def\L{{\cal L}}
\newcommand{\e}{{\rm e}}
\newcommand{\ep}{\epsilon}
\newcommand{\p}{\partial}
\newcommand{\vev}[1]{\left\langle #1 \right\rangle}
\newcommand{\psibar}{\overline{\psi}}
\newcommand{\cbar}{\bar{c}}
\newcommand{\dslash}{\hbox{$\partial$\kern-0.5em\raise0.3ex\hbox{/}}}
\def\slash#1{\hbox{$#1$\kern-0.5em\raise0.3ex\hbox{/}}}
\begin{document}
\begin{titlepage}
\rightline{KOBE-TH-00-05}
\rightline{hep-th/0008158}
\vspace{.5cm}
\begin{center}
{\LARGE On the gauge parameter dependence of QED}\\
\renewcommand{\thefootnote}{\fnsymbol{footnote}}
\vspace{1cm} Hidenori SONODA\footnote[2]{E-mail: {\tt
sonoda@phys.sci.kobe-u.ac.jp}}\\
\renewcommand{\thefootnote}{\arabic{footnote}}
\vspace{.2cm}
Physics Department, Kobe University, Kobe 657-8501, Japan\\
\vspace{.2cm} 
November 2000\\
\vspace{.2cm}
PACS numbers: 11.10.Gh, 11.15.-q\\
Keywords: renormalization, gauge field theories
\end{center}
\vspace{.3cm}
\begin{abstract}
The gauge parameter dependence of QED in the covariant gauge was given
explicitly long time ago by Landau and Khalatnikov.  We elucidate their
result by giving two new derivations.  The first derivation uses the
BRST invariance of the theory with a St\"uckelberg field, which is a
non-interacting fictitious Goldstone boson field.  The second derivation
is more straightforward but calculational.
\end{abstract}
\end{titlepage}

The purpose of this paper is to introduce a very simple trick for
determining the dependence of abelian gauge theories on the gauge fixing
parameter.  It is true that important physics lies only in the gauge
independent part of the theory, and anything dependent on the gauge
fixing parameter is unphysical.  Nevertheless, perturbation theory of
the manifestly renormalizable gauge theories cannot be formulated in a
gauge invariant way, and it is important to have a total control over
the gauge dependence of the correlation functions.

The gauge dependence of QED was discussed long ago by Landau and
Khalatnikov \cite{LKh}.  The familiar lagrangian of QED with electrons
is given by
\be
\L = {1 \over 4} F_{\mu\nu}^2 + {1 \over 2 \xi} \left( \p_\mu A_\mu
\right)^2 + \psibar \left( {1 \over i} \dslash - e \slash{A} + i M
\right) \psi
\ee
where $\xi$ is a gauge fixing parameter.\footnote{We will use the
euclidean metric throughout the paper.  The action $S$ is the integral
of the lagrangian density $\L$ over 4-dimensional euclidean space, and
the weight of functional integration is given by $\e^{-S}$.}  Landau and
Khalatnikov stated that the correlation functions in two different
gauges $\xi, \xi'$ are related by\footnote{Actually they gave a result
applicable to any gauge fixing function of $\partial_\mu A_\mu$.}
\be
\vev{A_{\mu_1} ... ~\psi ... ~\psibar ...}_{\xi'} =
\vev{(A_{\mu_1}+ \partial_{\mu_1} \phi) ... ~\e^{i e \phi} \psi
... ~\e^{- i e \phi} \psibar  ...}_\xi \label{relation}
\ee
where $\phi$ is a free real scalar field whose propagator is given by
\be
\vev{\phi (x) \phi (0)} \equiv (\xi'-\xi) \int_k {\e^{i k x} \over
(k^2)^2} \label{masslessphi}
\ee
This implies physically the free field nature of the longitudinal mode
of the photon.  Landau and Khalatnikov justified their result by showing
its consistency with the Ward identities.  The specific cases for the
two- and three-point functions were also verified by the method of
generating functionals in refs.~\cite{BSh,Z}.\footnote{In
ref.~\cite{BSh}, only the gauge parameter dependence of the electron
propagator was given explicitly although the arguments there can be
generalized.  In ref.~\cite{Z} the change of the generating functional
under an arbitrary infinitesimal change of the gauge fixing function was
given, and Eq.~(\ref{relation}), with $\phi$ contracted, was obtained
explicitly for the electron propagator and the vertex function.}

The aim of this paper is to rederive Eq.~(\ref{relation}) in an
illuminating way.  By introducing a St\"uckelberg field, we will derive
Eq.~(\ref{relation}) as a simple consequence of the BRST invariance.  We
will first prove the generalization of Eq.~(\ref{relation}) for the
massive QED.  Then, the result for QED with massless photons can be
obtained by taking the massless limit.

Let us consider the following lagrangian for the massive QED:
\ba
\L &=& {1 \over 4} F_{\mu\nu}^2 + {1 \over 2} \left( \p_\mu \varphi - m
A_\mu \right)^2 + {1 \over 2 \xi} \left( \p_\mu A_\mu - \xi m \varphi
\right)^2 \nonumber\\ && + \psibar \left( {1 \over i} \dslash - e
\slash{A} + i M \right) \psi + \p_\mu \bar{c} \p_\mu c + \xi m^2 \bar{c}
c \nonumber\\ &=& {1 \over 4} F_{\mu\nu}^2 + {1 \over 2 \xi} (\p \cdot
A)^2 + {m^2 \over 2} A_\mu^2 + {1 \over 2} \left( \left(\p_\mu
\varphi\right)^2 + \xi m^2 \varphi^2 \right) \nonumber \\&& + \psibar
\left( {1 \over i} \dslash - e \slash{A} + i M \right) \psi + \p_\mu
\bar{c} \p_\mu c + \xi m^2 \bar{c} c
\label{action}
\ea
The real scalar field $\varphi$, known as the St\"uckelberg field, is
the Goldstone boson field which gives a mass $m$ to the photon.  The
exponentiated field $\e^{i {e\over m} \varphi}$ is a charged scalar with
its magnitude frozen.  We have chosen the $R_\xi$ gauge so that
$\varphi$ becomes a free massive field of squared mass $\xi m^2$.
Ignoring the $\varphi$ and the anticommuting Faddeev-Popov (FP) ghost
fields $c, \bar{c}$ (or equivalently integrating them out), the
lagrangian reduces to the standard lagrangian for the massive QED with
electrons in the covariant gauge.

The lagrangian (\ref{action}) is invariant under the following BRST
transformation:
\ba
\delta_\ep A_\mu &=& \ep \p_\mu c, \quad \delta_\ep \varphi = m
\ep c\nonumber\\ \delta_\ep \psi &=& i e \ep c \psi, \quad \delta_\ep
\psibar = - i e \ep c \psibar\nonumber\\ \delta_\ep c &=& 0, \quad
\delta_\ep \cbar = \ep {1 \over \xi} \left( \p \cdot A - \xi m \varphi
\right)
\ea
where $\ep$ is an arbitrary anticommuting constant.  Out of the four
fields $A_\mu, \psi, \psibar$, and $\varphi$ we can construct three
gauge and BRST invariant fields:
\ba
A'_\mu &\equiv& A_\mu - {1 \over m} \p_\mu \varphi \label{BRSTinvA}\\
\psi' &\equiv& \e^{- i {e \over m} \varphi} \psi, \quad
\psibar' \equiv \e^{i {e \over m} \varphi} \psibar\label{BRSTinvpsi}
\ea
It is easy to compute the $\xi$ dependence of the lagrangian:
\ba
{\p \L \over \p \xi} &=& - {1 \over 2 \xi^2} (\p \cdot A)^2 + {1 \over
2} m^2 \varphi^2 + m^2 \bar{c} c \nonumber\\ &=& - {1 \over 2 \xi}{1
\over \ep} \delta_\ep \left[ \bar{c} \left( \p \cdot A + \xi m \varphi
\right) \right] + {1 \over 2 \xi} \bar{c} ( - \p^2 + \xi m^2) c
\ea
This implies that the correlations of BRST invariant fields that are
independent of the FP ghosts do not depend on the gauge fixing parameter
$\xi$.  Therefore, the correlation functions of $A'_\mu, \psi',
\psibar'$ are independent of $\xi$:
\be
\vev{A'_{\mu_1} ... \psi' ... \psibar' ... }_{\xi'}
= \vev{A'_{\mu_1} ... \psi' ... \psibar' ... }_{\xi} \label{BRSTinv}
\ee
The explicit gauge dependence of the correlation functions can be
obtained by contracting the free scalar $\varphi$ in this formula.

We now wish to derive the Landau-Khalatnikov result (\ref{relation}) for
the massive QED:
\be
\vev{A_{\mu_1} ... \psi ... \psibar ...}_{\xi'} = \vev{\left(A_{\mu_1} +
\p_{\mu_1} \phi \right) ... \psi \e^{i e \phi}
... \psibar \e^{- ie \phi} ...}_\xi \label{massive_relation}
\ee
where the propagator of the free scalar field $\phi$ is given by
\ba
\vev{\phi (x) \phi (0)} &\equiv& {1 \over m^2} \int_k \e^{i k x} \left(
{1 \over k^2 + \xi m^2} - {1 \over k^2 + \xi' m^2} \right)\nonumber\\
&=& (\xi'-\xi) \int_k \e^{i k x} {1 \over (k^2+\xi m^2)(k^2 + \xi'
m^2)} \label{massivephi}
\ea
Instead of proving Eq.~(\ref{massive_relation}) directly, we prove the
following equivalent relation that is obtained by substituting
Eq.~(\ref{massive_relation}) into Eq.~(\ref{BRSTinv}):
\ba
&&\vev{\left( A_{\mu_1} + \p_{\mu_1} \left( \phi - {1 \over m}
\varphi_{\xi'} \right)\right) ... \psi \e^{i e \left(\phi - {1 \over m}
\varphi_{\xi'}\right)} ... \psibar \e^{- i e \left(\phi - {1 \over m}
\varphi_{\xi'}\right)} ...}_\xi\nonumber\\ &=& \vev{\left( A_{\mu_1} -
\p_{\mu_1} {1 \over m} \varphi_\xi \right) ... \psi \e^{- i {e \over m}
\varphi_\xi} ... \psibar \e^{i {e \over m} \varphi_\xi} ...}_\xi
\label{toshow}
\ea
where $\varphi_\xi$ is the St\"uckelberg field for the gauge fixing
parameter $\xi$ with the propagator
\be
\vev{\varphi_\xi (x) \varphi_\xi (0)} = \Delta (x-y; \xi m^2) \equiv
\int_k {\e^{i k x} \over k^2 + \xi m^2} \label{delta}
\ee
Eq.~(\ref{toshow}) is valid if and only if
\be
\vev{\e^{ie \left( \phi - {1 \over m} \varphi_{\xi'}\right)} ...
\e^{- ie \left(\phi - {1 \over m} \varphi_{\xi'}\right)} ...}
= \vev{ \e^{- i {e \over m} \varphi_\xi} ... \e^{i {e \over m}
\varphi_\xi} ...}
\ee
But this is equivalent to
\be
\vev{\phi \phi} = {1 \over m^2} \left( \vev{\varphi_\xi \varphi_\xi} -
\vev{\varphi_{\xi'} \varphi_{\xi'}}\right), \label{phi}
\ee
which is precisely the definition of the propagator of $\phi$ given in
Eq.~(\ref{massivephi}).  This concludes the proof of
Eq.~(\ref{massive_relation}).\footnote{Eq.~(12.9.21) of ref.~\cite{C1}
gives the $\xi$ derivative of arbitrary correlation functions.  By
integrating the equation over $\xi$, Eqs.~(\ref{massive_relation}) of
the main text can be obtained in principle.}  The propagator of $\phi$
is the difference of the propagators of the St\"uckelberg fields in two
gauges.

The original result (\ref{relation}) for QED with massless photons can
be derived from Eq.~(\ref{massive_relation}) simply by taking the limit
$m \to 0$; in the limit we get the massless propagator
(\ref{masslessphi}) from the massive propagator (\ref{massivephi}).
However, the proof itself does not go through for $m=0$.  With $m=0$ the
BRST invariant lagrangian (\ref{action}) still makes sense, but the
definitions (\ref{BRSTinvA},\ref{BRSTinvpsi}) suffer from the diverging
factor ${1 \over m} \to \infty$.  A slight modification is necessary for
the proof.  We introduce the following lagrangian:
\be
\L = {1 \over 4} F^2 + {1 \over 2 \xi} (\p \cdot A)^2 - {1 \over 2 \xi}
\left( \p^2 \varphi \right)^2 + \psibar \left( {1 \over i} \dslash - e
\slash{A} + i M \right) \psi + \p_\mu \bar{c} \p_\mu c
\ee
This is invariant under the BRST transformation:
\ba
\delta_\ep A_\mu &=& \ep \p_\mu c,\quad \delta_\ep \varphi = \ep
c\nonumber\\ \delta_\ep \psi &=& i e \ep c \psi, \quad \delta_\ep
\psibar = - i e \ep c \psibar\nonumber\\ \delta_\ep c &=& 0, \quad
\delta_\ep \cbar = \ep {1 \over \xi} \left( \p \cdot A - \p^2 \varphi
\right)
\ea
This transformation is strictly nilpotent.  From the BRST invariance of
the redefined fields
\ba
A_\mu' &\equiv& A_\mu - \p_\mu \varphi\nonumber\\
\psi' &\equiv& \e^{- i e \varphi} \psi, \quad
\psibar' \equiv \e^{i e \varphi} \psibar,
\ea
the rest of the proof follows exactly the same way.

There is a more straightforward way of deriving the gauge parameter
dependence (\ref{massive_relation}).  Before proceeding with the
derivation, however, let us work out a few concrete consequences of the
Landau-Khalatnikov equation (\ref{massive_relation}) for completeness of
the paper.  We first consider the electron propagator.
Eq.~(\ref{massive_relation}) implies
\be
\vev{\psi (x) \psibar (y)}_\xi = \exp \left[ {\xi e^2 \over (4\pi)^2}
\left( - {\cal D} (0;\xi m^2) + {\cal D} (x-y; \xi m^2)\right) \right]
\vev{\psi (x) \psibar (y)}_{\xi = 0} \label{bare_prop}
\ee
where
\be
{1 \over (4\pi)^2}{\cal D}(x;\xi m^2) \equiv {1 \over \xi m^2}
\left( \Delta (x; 0) - \Delta (x;\xi m^2)\right) = \int_k {\e^{i k
x} \over k^2(k^2+\xi m^2)} \label{calD}
\ee
The value of ${\cal D} (0;\xi m^2)$ is ultraviolet (UV) divergent, and
in the dimensional regularization it is calculated as
\be
{1 \over (4\pi)^2} {\cal D} (0;\xi m^2) = \mu^\ep \int {d^D k \over
(2 \pi)^D} { 1 \over k^2(k^2 + \xi m^2)} = {1 \over (4 \pi)^2} \left( {2
\over \ep} + 1 - \ln {\xi m^2 \over \bar{\mu}^2} \right)
\label{calD_zero}
\ee
where $D \equiv 4 - \ep$, and $\bar{\mu}^2 \equiv 4 \pi \mu^2 \e^{-
\gamma}$ ($\gamma$ is the Euler constant).  Let us introduce
renormalization constants as follows:
\ba
A_{\mu, r} &\equiv& {1 \over \sqrt{Z_3}} A_\mu, \quad
\psi_r \equiv {1 \over \sqrt{Z_2 (\xi)}} \psi \nonumber\\
e_r^2 &\equiv& Z_3 e^2, \quad m_r^2 \equiv Z_3 m^2, \quad
\xi_r \equiv {1 \over Z_3} \xi
\ea
where $Z_3$ is independent of $\xi$.  Eqs.~(\ref{bare_prop}, \ref{calD},
\ref{calD_zero}) imply that, in the minimal subtraction (MS) scheme, the
wave function renormalization constant depends on the gauge fixing
parameter as
\be
Z_2 (\xi) = \exp \left[ - ~{\xi_r e_r^2 \over (4 \pi)^2} {2 \over \ep}
\right] Z_2 (0).
\ee
This relation was first obtained by Johnson and Zumino
\cite{JZ}.\footnote{A large regulator mass was used as a UV cutoff in
\cite{JZ}.  The calculation with the dimensional regularization was
first made by Collins \cite{C1,C2} and Lautrup \cite{L}.}  After
renormalization, Eq.~(\ref{bare_prop}) gives
\cite{LKh,BSh,Z}\footnote{See also ref.~\cite{B} for a derivation using
the functional method.}
\ba
&&\vev{\psi_r (y) \psibar_r (z)}_{\xi_r}\Big/\vev{\psi_r (y) \psibar_r
(z)}_{\xi_r = 0} \nonumber\\ &=& \exp \left[ {\xi_r e_r^2 \over
(4\pi)^2} \left( \ln {\xi_r m_r^2 \over \bar{\mu}^2} - 1 + {\cal D}
(y-z; \xi_r m_r^2) \right) \right]
\ea
where ${\cal D}$, which is UV finite for non-vanishing arguments, is
defined by Eq.~(\ref{calD}).

We next consider the three-point function.  Eq.~(\ref{massive_relation})
gives, after renormalization,
\ba
&&\vev{A_{\mu,r} (x) \psi_r (y) \psibar_r (z)}_{\xi_r} = \exp \left[
{\xi_r e_r^2 \over (4\pi)^2} \left( \ln {\xi_r m_r^2 \over \bar{\mu}^2}
- 1 + {\cal D} (y-z; \xi_r m_r^2) \right) \right] \nonumber\\ && \times
\Bigg[ ~\vev{A_{\mu,r} (x) \psi_r (y) \psibar_r (z)}_0 \\
&&~+ ~{i \xi_r e_r \over (4\pi)^2} \left( \p_\mu {\cal D} (x-y;\xi_r
m_r^2) - \p_\mu {\cal D} (x-z;\xi_r m_r^2)\right) \vev{\psi_r (y)
\psibar_r (z)}_0 ~\Bigg]\nonumber
\ea
This was obtained in refs.~\cite{LKh,Z} for $m_r^2 = 0$.

For arbitrary correlation functions in the MS scheme, we find
\ba
&&\vev{A''_{\mu, r} ... \psi_r (y_1) ... \psi_r (y_F) \psibar_r (z_1)
...  \psibar_r (z_F) }_{\xi_r}\Big/ \vev{A_{\mu, r} ... \psi_r
... \psibar_r ...}_0 \nonumber\\ &=& \exp \Bigg[ ~{\xi_r e_r^2 \over
(4\pi)^2} \Bigg\lbrace F \left( \ln {\xi_r m_r^2 \over \bar{\mu}^2} -
1\right) + \sum_{i,j} {\cal D} (y_i - z_j; \xi_r m_r^2)\nonumber\\ &&
\quad - \sum_{i<j} \left( {\cal D} (y_i - y_j; \xi_r m_r^2) + {\cal D}
(z_i - z_j; \xi_r m_r^2) \right) \Bigg\rbrace ~\Bigg]
\label{renormalized}
\ea
where only the transverse part
\be
A''_\mu \equiv A_\mu - \p_\mu {1 \over \p^2} \p \cdot A
\ee
has been considered for simplicity.  Eq.~(\ref{renormalized}) implies
that the correlation functions of the elementary fields at large
distances are independent of the gauge fixing parameter.  This would
imply the gauge independence of the S-matrix elements in the Minkowski
space.

Now, let us proceed with the second derivation of
Eq.~(\ref{massive_relation}).  We start with the following lagrangian
for the massive QED:
\ba
\L &=& {1 \over 4} F_{\mu\nu}^2 + {1 \over 2} m^2 A_\mu^2 + {1 \over 2
\xi} (\p \cdot A)^2 + \psibar \left( {1 \over i} \dslash - e \slash{A} +
i M \right) \psi \nonumber\\
&& \quad + {1 \over 2 (\xi' - \xi)}~\phi (- \p^2 + \xi m^2)(- \p^2 +
\xi' m^2) \phi \label{first}
\ea
where we have added a non-interacting scalar field $\phi$, which is the
same field that appears in Eq.~(\ref{massive_relation}).  Since $\phi$
is decoupled, we can integrate it out to get the standard lagrangian for
the massive QED with the gauge fixing parameter $\xi$.  We introduce a
change of variables
\be
A'_\mu = A_\mu + \p_\mu \phi,\quad \psi' = \e^{i e \phi} \psi, \quad
\psibar' = \e^{- i e \phi} \psibar
\ee
This is nothing but a gauge transformation with the gauge function
$\phi$.  Note that the right-hand side of Eq.~(\ref{massive_relation})
is the correlation of $A'_\mu, \psi', \psibar'$ using the above
lagrangian (\ref{first}).  In terms of the redefined fields, we can
rewrite the lagrangian as
\ba
&& \L = {1 \over 4} {F'_{\mu\nu} }^2 + {1 \over 2} m^2 \left( A'_\mu -
\p_\mu \phi \right)^2 + {1 \over 2 \xi} \left( \p \cdot A' - \p^2 \phi
\right)^2 \nonumber\\ && ~+ \psibar' \left( {1 \over i} \dslash - e
\slash{A}' + i M \right) \psi' + {1 \over 2(\xi'-\xi)}~\phi (- \p^2 +
\xi m^2)(- \p^2 + \xi' m^2) \phi \nonumber\\ &=& {1 \over 4}
{F'_{\mu\nu} }^2 + {1 \over 2} m^2 {A'}^2 + {1 \over 2 \xi} (\p \cdot
A')^2 + \overline{\psi'} \left( {1 \over i} \dslash - e \slash{A}' + i M
\right) \psi' \nonumber\\ && + {1 \over \xi} \phi \left( - \p^2 + \xi
m^2 \right) \p \cdot A' + {\xi' \over 2 \xi (\xi'-\xi)} \phi \left( -
\p^2 + \xi m^2 \right)^2 \phi
\ea
Integrating out $\phi$, we get
\be
\L' = {1 \over 4} {F'_{\mu\nu} }^2 + {1 \over 2} m^2 {A'}^2 + {1 \over 2
\xi'} (\p \cdot A')^2 + \overline{\psi'} \left( {1 \over i} \dslash 
- e \slash{A}' + i M \right) \psi', \label{second}
\ee
which is a lagrangian with a new gauge fixing parameter $\xi'$.  This
implies that the correlation of $A_\mu', \psi', \psibar'$ using the
lagrangian (\ref{first}) is the same as the correlation of the same
fields using the lagrangian (\ref{second}).  Hence, we obtain
Eq.~(\ref{massive_relation}).

We wish to apply the techniques developed above to the gauge dependence
of the abelian Higgs theory both in the covariant gauge and in the
$R_\xi$ gauge.  In the covariant gauge the theory is defined by the
lagrangian
\be
\L = {1 \over 4} F_{\mu\nu}^2 + {1 \over 2 \xi} (\p \cdot A)^2 +
|(\p_\mu - i e A_\mu) \phi|^2 + M^2 |\phi|^2 + {\lambda \over 4}
|\phi|^4 
\ee
The $\xi$ dependence of the correlation functions of $A_\mu, \phi,
\phi^*$ can be obtained in the same way as for the QED with electrons.
Let us consider the implications of the $\xi$ dependence thus obtained.
In the Higgs phase the correlation function
\be
\vev{A_\mu ... \phi (y_1) ... \phi (y_B) \phi^* (z_1) ... \phi^*
(z_{\bar{B}})} \nonumber
\ee
is non-vanishing even if $B \ne \bar{B}$.  The correlation functions for
$B \ne \bar{B}$ have $\xi$ dependent infrared (IR) divergences.  For
example, we find, after UV renormalization, that
\be
\vev{\phi_r}_{\xi_r} = \e^{{\xi_r e_r^2 \over (4 \pi)^2} {1 \over \ep}}
\vev{\phi_r}_0
\ee
For the two-point functions, we obtain
\ba
2 \vev{\Re \phi_r (x) \Re \phi_r (y) }_{\xi_r} &=& \left( \pi \e^\gamma
\mu^2 (x-y)^2\right)^{- {\xi_r e_r^2 \over (4\pi)^2}} \vev{\phi (x)
\phi^* (y)}_0\\ && + \e^{{\xi_r e_r^2 \over (4\pi)^2} {4 \over \ep}}
\left( \pi \e^\gamma \mu^2 (x-y)^2\right)^{{\xi_r e_r^2 \over (4\pi)^2}}
\vev{\phi (x) \phi (y)}_0 \nonumber\\ 2 \vev{\Im \phi_r (x) \Im \phi_r
(y)}_{\xi_r} &=& \left( \pi \e^\gamma \mu^2 (x-y)^2\right)^{- {\xi_r
e_r^2 \over (4\pi)^2}} \vev{\phi (x) \phi^* (y)}_0 \\ && - \e^{{\xi_r
e_r^2 \over (4\pi)^2} {4 \over \ep}} \left( \pi \e^\gamma \mu^2
(x-y)^2\right)^{{\xi_r e_r^2 \over (4\pi)^2}} \vev{\phi (x) \phi
(y)}_0\nonumber
\ea
The second terms on the right-hand sides are IR divergent for
non-vanishing $\xi_r$.  All IR divergences in the covariant gauge depend
on $\xi_r$, and they can be determined explicitly.

Finally, we consider the Higgs theory in the $R_\xi$ gauge \cite{FLS}.
In this case the fictitious Goldstone boson does not decouple for the
massive photon, and we cannot obtain the $\xi$ dependence of the
correlations of elementary fields.  The lagrangian in the $R_\xi$ gauge
is given by
\ba
&&\L_{R_\xi} = {1 \over 4} F_{\mu\nu}^2 + {1 \over 2 \xi} \left(\p \cdot
A - \xi e v \chi \right)^2 \nonumber\\ &&\quad + |(\p_\mu - i e A_\mu)
\phi|^2 + M^2 |\phi|^2 + {\lambda \over 4} |\phi|^4 \label{Rxi} + \p_\mu
\bar{c} \p_\mu c + \xi e^2 v \rho \bar{c} c
\ea
where $\rho, \chi$ are the real and imaginary parts of $\phi$:
\be
\phi = {1 \over \sqrt{2}} \left( \rho + i \chi \right)
\ee
The mass parameter $v$ has been introduced to remove the tree-level
mixing between $\chi$ and $\p \cdot A$.  A St\"uckelberg field can be
introduced as
\ba
\L &=& {1 \over 4} F_{\mu\nu}^2 + {1 \over 2} \left( \p_\mu \varphi - m
A_\mu \right)^2 + {1 \over 2 \xi} \left(\p \cdot A - \xi e v \chi - \xi
m \varphi \right)^2 \nonumber\\ &&\quad + |(\p_\mu - i e A_\mu) \phi|^2
+ M^2 |\phi|^2 + {\lambda \over 4} |\phi|^4\label{Rxiphi} \\ &&\quad +
\p_\mu \bar{c} \p_\mu c + \xi \left( e^2 v \rho + m^2 \right)
\bar{c} c\nonumber
\ea
The $\varphi$ field is not free anymore; it couples to $\chi$ through
the term $\xi e v m \varphi \chi$.  Regarding the field $\chi$ as a
source, $\varphi$ can be integrated out.  In the massless limit $m \to
0$, the field $\varphi$ decouples from $\chi$, and the lagrangian
(\ref{Rxiphi}) reduces to (\ref{Rxi}).  But, as we will see shortly, the
effect of the coupling of $\varphi$ with $\chi$ remains even in the
massless limit.

The lagrangian (\ref{Rxiphi}) is invariant under the following BRST 
transformation:
\ba
\delta_\ep A_\mu &=& \ep \p_\mu c,\quad \delta_\ep \varphi = m
\ep c\nonumber\\ \delta_\ep \rho &=& - e \ep c \chi, \quad \delta_\ep
\chi = e \ep c \rho \nonumber\\ \delta_\ep c &=& 0, \quad \delta_\ep
\cbar = \ep {1 \over \xi} \left( \p \cdot A - \xi ev\chi - \xi m \varphi
\right)
\ea
and the BRST invariant fields are defined by
\ba
A''_\mu &\equiv& A_\mu - \p_\mu {1 \over \p^2} \p \cdot A\\
\phi' &\equiv& \e^{ - i {e \over m} \varphi} \phi, \quad
{\phi'}^* \equiv \e^{i {e \over m} \varphi} \phi^*
\ea
Using the $\xi$ independence of the correlation of $A''_\mu, \phi',
{\phi'}^*$, we obtain
\ba
&& \Bigg\langle A''_\mu ... \phi (y_1) ... \phi (y_B) \phi^* (z_1) ...
\phi^* (z_{\bar{B}})\nonumber\\ &&\quad \times \exp \Bigg[ {1 \over 2}
\xi^2 e^2 v^2 m^2 \int_{r,r'} \chi (r) \Delta(r-r';\xi m^2) \chi (r')
\nonumber\\ && + ~i \xi e^2 v \int_r \chi (r) \bigg( \sum_{1\le i\le B}
\Delta (r-y_i; \xi m^2) - \sum_{1\le i\le \bar{B}} \Delta (r-z_i; \xi
m^2) \bigg) \Bigg] \Bigg\rangle_\xi\nonumber \\&=& \exp \Bigg[ {\xi e^2
\over (4\pi)^2} \bigg( - {1 \over 2} (B+ \bar{B}) {\cal D} (0; \xi m^2)
+ \sum_{i\le B,j \le \bar{B}} {\cal D} (y_i - z_j; \xi m^2)\nonumber\\
&&\quad - \sum_{i < j \le B} {\cal D} (y_i - y_j; \xi m^2) +\sum_{i < j
\le \bar{B}} {\cal D} (z_i - z_j; \xi m^2) \bigg)\Bigg]\nonumber\\
&&\qquad \times \vev{A_\mu ... \phi (y_1) ... \phi (y_B) \phi^* (z_1)
...  \phi^* (z_{\bar{B}})}_0
\ea
where the correlation is evaluated with the lagrangian (\ref{Rxi}) in
the $R_\xi$ gauge.  We observe that on the left-hand side, the source
terms quadratic in $\chi$ vanish in the limit $m \to 0$, but not the
terms linear in $\chi$.  Hence, even in the massless limit, the above
formula does not give the $\xi$ dependence of the correlation of
elementary fields alone.  In the limit $m \to 0$, we obtain the
following $\xi$ dependence after renormalization in the MS scheme:
\ba
&& \Bigg\langle A''_{\mu,r} ... \phi_r (y_1) ... \phi_r^* (z_1) ...
\nonumber\\ && \times \exp \Bigg[ ~i \xi_r e_r^2 v_r \int_r \chi_r
(r) \bigg( \sum_{1 \le i \le B} \Delta (y_i - r;0) - \sum_{1 \le i \le
\bar{B}} \Delta (z_i - r;0) \bigg) \Bigg] \Bigg\rangle_{\xi_r}
\nonumber\\ &=& \e^{ {\xi_r e_r^2 \over (4 \pi)^2} {1 \over \ep}
(B-\bar{B})^2} \times \Pi_{i,j} \left ( \pi \e^\gamma \mu^2
(y_i-z_j)^2\right)^{- {\xi_r e_r^2 \over (4\pi)^2}} \nonumber\\ &&\times
~ \Pi_{i<j\le B} \left( \pi \e^\gamma \mu^2 (y_i-y_j)^2\right)^{{\xi_r
e_r^2 \over (4\pi)^2}} \times \Pi_{i<j\le \bar{B}} \left( \pi \e^\gamma
\mu^2 (z_i-z_j)^2\right)^{{\xi_r e_r^2 \over (4\pi)^2}} \nonumber\\&&
\times ~ \vev{A_{\mu,r} ... \phi_r (y_1) ... \phi_r^* (z_1) ...}_0
\ea
Note that the IR divergence in the first factor on the right-hand side
is due to the long-range source term coupled to $\chi_r$.\footnote{In
the $R_\xi$ gauge there is no IR divergence in the correlations of
elementary fields.  Here the IR divergences arise due to the long range
function $\Delta(x;0) = {\rm O} (1/x^2)$ in the source term.}

\vspace{.2cm} In this paper we have given two new proofs of the result
of Landau and Khalatnikov on the exact gauge dependence of the
correlation functions of the elementary fields for the abelian gauge
theories in the covariant gauge.  The trick of the St\"uckelberg field
can be introduced also to the non-abelian gauge theories, but due to the
coupling of the St\"uckelberg field to the gauge and FP ghost fields, no
simple formulas can be obtained.

\vspace{.2cm} I thank the referee for an appropriate suggestion for
revision.  The first version of the paper was written during a visit to
the particle theory group of Hokkaido University.  I would like to thank
Prof.~Noboru Kawamoto and the other members of the group for
hospitality.  This work was supported in part by the Grant-In-Aid for
Scientific Research from the Ministry of Education, Science, and
Culture, Japan (\#11640279).


\begin{thebibliography}{9}
\bibitem{LKh} L.~D.~Landau, I.~M.~Khalatnikov, ``The Gauge
Transformation of the Green's Function for Charged Particles,'' JETP
{\bf 2}(1956)69--72
\bibitem{BSh} A.~V.~Svidzinskii, Thesis, L'vov State University, 1955;\\
N.~N.~Bogoliubov, D.~V.~Shirkov, {\it Introduction to the Theory of
Quantized Fields} (Wiley-Interscience, 1959) Chapter VII, section 40
\bibitem{Z} B.~Zumino, ``Gauge Properties of Propagators in Quantum
Electrodynamics,'' J.~Math.~Phys. {\bf 1}(1960)1--7
\bibitem{C1} J.~C.~Collins, {\it Renormalization} (Cambridge University
Press, 1984) Chapter 12.9
\bibitem{JZ} K.~Johnson, B.~Zumino, ``Gauge Dependence of the
Wave-Function Renormalization Constant in Quantum
Electrodynamics,''\hfill\break Phys.~Rev.~Lett. {\bf 3}(1959)351--352
\bibitem{C2} J.~C.~Collins, Ph.D. thesis, Cambridge University, 1975
\bibitem{L} B.~Lautrup, ``Renormalization Constants and Asymptotic
Behavior in Quantum Electrodynamics,'' Nucl. Phys. {\bf
B105}(1976)23--44
\bibitem{B} L.~S.~Brown, {\it Quantum Field Theory} (Cambridge
University Press, 1992) Chapter 8.4
\bibitem{FLS} K.~Fujikawa, B.~W.~Lee, A.~I.~Sanda, ``Generalized
Renormalizable Gauge Formulation of Spontaneously Broken Gauge
Theories,'' Phys.~Rev. {\bf D6}(1972)2923--2943
\end{thebibliography}
\end{document}